\documentclass{appolb}
\usepackage{graphicx}

\begin{document}
\title{Future Collider Measurements for Cosmic Ray Induced Air Shower Modelling
\thanks{Presented at ``Diffraction and Low-$x$ 2024'', Trabia (Palermo, Italy), September 8-14, 2024.}%
}
\author{Clara E. Leitgeb\footnote{Corresponding author: \texttt{clara.leitgeb@physik.hu-berlin.de}}, Robert D. Parsons
\address{Humboldt-Universit\"at zu Berlin, Germany}
\\[3mm]
{Andrew Taylor 
\address{Deutsches Elektronensynchrotron DESY, Germany}
}
\\[3mm]
{Kenneth Ragan
\address{Mc.Gill University, Canada}
}
\\[3mm]
{David Berge and Cigdem Issever
\address{Humboldt-Universit\"at zu Berlin, Germany and DESY, Germany}
}
}

\maketitle
\begin{abstract}
The identification of gamma-ray induced air showers with Cherenkov telescopes suffers from contamination with a specific class of cosmic ray induced air showers. The predictions for this background show strong discrepancies between the available event generators. In this study, we identify collision events of cosmic rays with atmospheric nuclei in which a large fraction of the original beam energy is transmitted to the electromagnetic part of the shower as the main source for this background. Consequently, we define a pseudorapidity region of interest for hadron collider experiments that corresponds to this background, taking into account the center-of-mass energy. This region of interest is compared with the available datasets and the pseudorapidity coverage of the detectors that recorded it. We find that the LHCf and RHICf detectors are the only ones covering substantial parts of this region of interest and suggest a measurement of the energy spectra of reconstructed neutral pions to be made with this data. Such results could serve as valuable constraints for a future parameter tuning of the event generators to improve the background estimation uncertainties for gamma-ray induced air shower identification. 

\end{abstract}
  
\section{Introduction}

Imaging Atmospheric Cherenkov Telescopes (IACTs) \cite{2015arXiv151005675H} are used to detect Cherenkov light from air showers that were initiated by cosmic gamma-rays. The electromagnetic interaction between the gamma-rays and the atmospheric atoms can be accurately modelled with quantum electrodynamics. However, in the energy range that IACTs are sensitive to, gamma-ray initiated showers are outnumbered by a factor of roughly $10^{3}$ \cite{ParticleDataGroup:2024cfk} by showers caused by cosmic ray (CR) particles. 

CRs (for the energy range of interest here) consist predominantly of protons and span a wide range of energies. When they arrive at the Earth they collide with atmospheric nuclei, mostly Nitrogen and Oxygen. The collisions are dominated by strong interactions with low momentum exchange, so-called soft QCD processes. The resulting particles initiate a cascade of further particle production, also referred to as extensive air showers (EAS). The produced particle multiplicities, the shower shape and maximum depend strongly on the details of the initial collision of the CR and the atmospheric nucleus. However, the amplitudes of soft QCD processes cannot be derived via perturbative QCD. The event generators most commonly used to simulate these kind of processes, EPOS-LHC \cite{EPOSLHC}, QGSJET II-04 \cite{Ostapchenko_2011}, Sibyll 2.3d \cite{Sibyll23d} and Pythia 8.3 \cite{Pythia83} are strongly dependent on reference measurements from hadron colliders for the tuning of their free parameters. 

Usually, gamma-ray initiated EAS can be separated well from the CR background (at a level of $\sim 99\%$) due to their differences in the shower shapes: typically, CR induced EAS are wider and longer than gamma-ray initiated EAS \cite{2015arXiv151005675H}. However, despite the good background separation, the amount of remaining background events is still significant due to the much larger number of CR induced EAS compared to gamma-ray induced EAS.  

In this work, we investigate the main source of the CR background for IACT gamma-ray analyses and the modelling performance of the currently available event generators. We define a suitable phase space for hadron collider measurements corresponding to these events. Furthermore, we identify detectors at colliders covering that phase space to improve the modelling of the CR background for IACT gamma-ray analyses.

\section{IACT Backgrounds from Cosmic Ray Induced EAS}

\begin{figure}[htb]
\centerline{%
\includegraphics[scale=0.3]{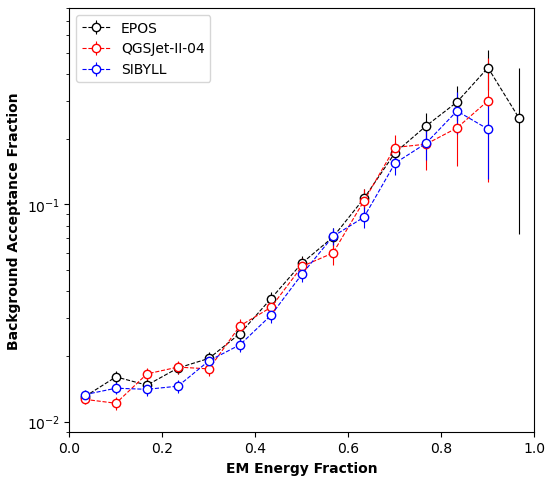}}
\caption{Fraction of EAS induced by a $1\,\rm{TeV}$ proton colliding with an Nitrogen atom passing gamma-ray selection criteria as a function of the fraction of energy carried by the EM parts of the shower. The individual event generators predict roughly the same behaviour.}
\label{Fig:BkgEj}
\end{figure}

The properties of CR backgrounds passing the gamma-ray identification selections are studied with simulations using CORSIKA-7 \cite{Heck_1998}. Collisions of cosmic protons with an energy of $1\,\rm{TeV}$ with atmospheric Nitrogen nuclei have been simulated at an altitude of $17\,\rm{km}$. The particle information was recorded both instantly after the primary interaction and at ground level. The simulated Cherenkov radiation was then propagated into a simple simulated Cherenkov telescope setup of nine telescopes placed in a $3\times 3$ grid with a distance of $120\,\rm{m}$ between the telescopes. The selection of gamma-ray like events is performed by applying cuts on the measured mean scaled width and the mean scaled length parameters of the showers. The fraction of events passing this selection is then compared to the fraction of the total shower energy that can be attributed to electromagnetic (EM) components of the shower, i.e. photons, electrons and neutral pions as shown in Fig.\,\ref{Fig:BkgEj}. It can be observed that the CR background increases with the EM energy fraction, with more that 20\% of the events passing the selection criteria at an EM energy fraction of 70\% of the initial proton's energy. This behaviour is predicted by all three shown event generators.

\begin{figure}[htb]
\centerline{%
\includegraphics[scale=0.4]{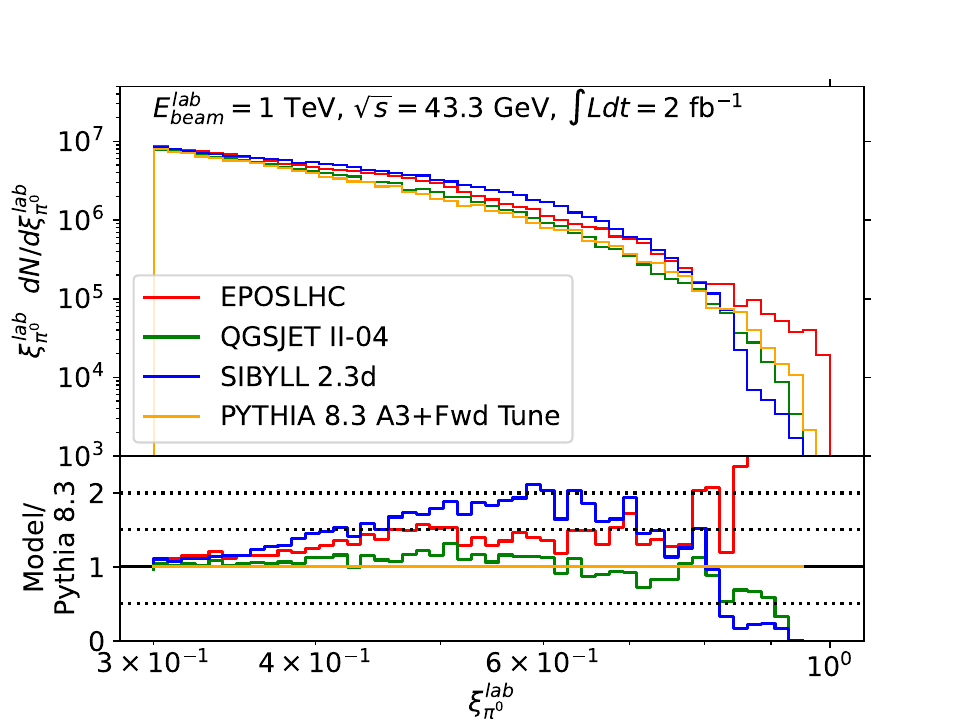}
\includegraphics[scale=0.4]{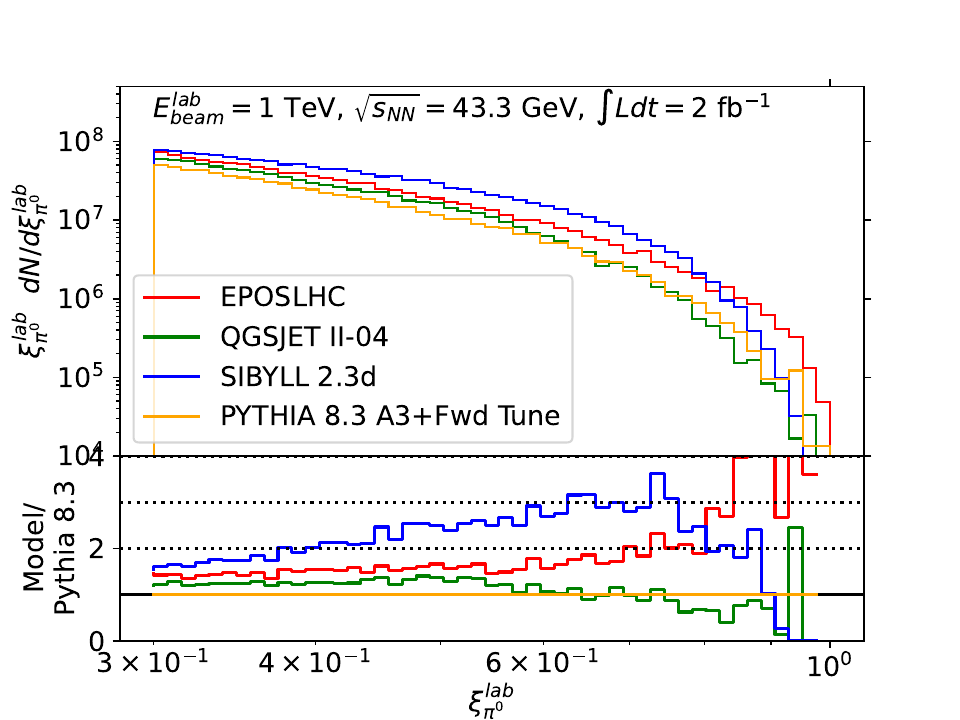}}
\caption{Energy fraction of neutral pions of the original $1\,\rm{TeV}$ proton beam, that collides with a fixed proton target (left) or an Oxygen target (right). The colored lines correspond to four different event generators.}
\label{Fig:PionE1TeV}
\end{figure}

Large energy fractions in the EM component of the shower are an indicator of the production of a highly energetic neutral pion which is already produced at an early shower stage. The left plot of Fig.\,\ref{Fig:PionE1TeV} shows the predicted energy distribution of neutral pions in a collision of a $1\,\rm{TeV}$ proton with a proton at rest. The $x$-axis depicts the fraction of the pion's energy of the original proton beam energy. The plot focuses on the high-energy end of the spectrum for the four different event generators. This demonstrates that the event generators differ strongly in their predictions for the process class that populates the CR EAS background for IACT analyses.

\section{Pseudorapidity Region of Interest and Scaling Relation}

In order to define the phase space relevant for CR background events, we adopt 70\% as a threshold for the energy carried by the neutral pion in comparison to the original beam energy. This translates into pseudorapidity boundaries that depend on the center-of-mass energy $\sqrt{s}$:\\
\begin{equation}
\eta_{\rm{max}} \approx \ln\bigg(\frac{\sqrt{s}}{m_{\pi^{0}}}\bigg),\hspace{1cm}
\eta_{\rm{min}} \approx \ln\bigg(\frac{0.7\sqrt{s}}{\sqrt{m_{\pi^{0}}^{2} + p_{\rm{T},\pi^{0}}^{2}}}\bigg)\,\, .
\end{equation}

\begin{figure}[htb]
\centerline{%
\includegraphics[scale=0.5]{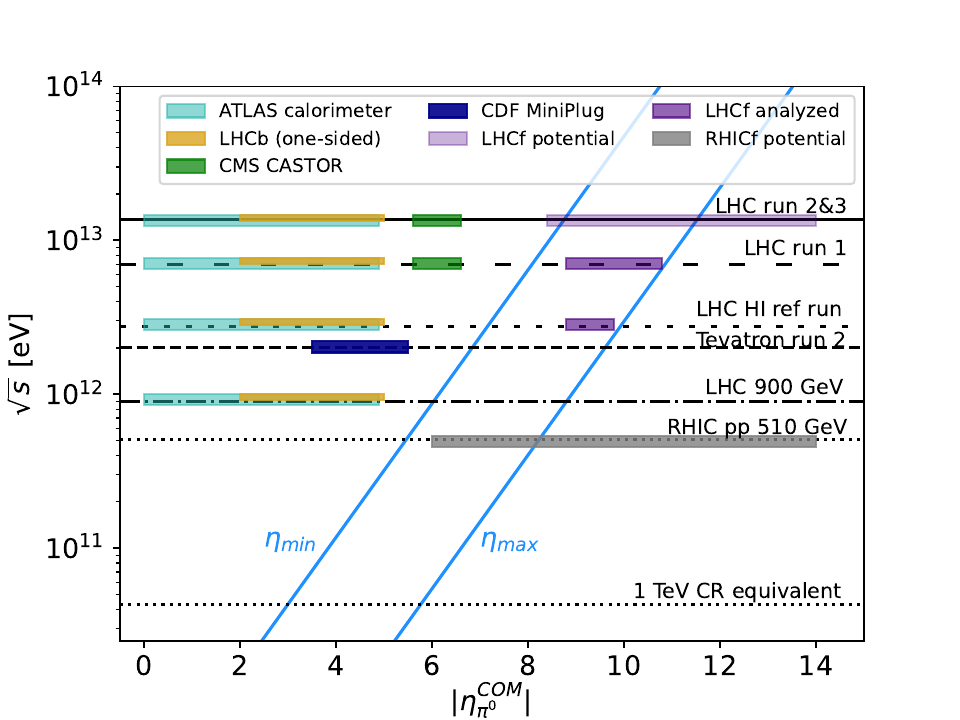}}
\caption{Region of interest for CR background events in the plane spanned by pseudoraptidity and $\sqrt{s}$. The horizontal lines show collider datasets at the respective center of mass energy. The colored boxes show the pseudorapidity coverage of individual detectors.}
\label{Fig:EtaRanges}
\end{figure}

Here, the lower boundary is not only defined by the 70\% requirement on the energy, but also by the sum of the neutral pion mass squared and the squared pion momentum in the transverse plane with respect to the beam axis. The latter is estimated from simulations: for different center-of-mass energies, the maximum $p_{\rm{T}}$ of the neutral pion does not change much and is typically found at $\sim 1.5\,\rm{GeV}$. The pseudorapidity range defined by $\eta_{\rm{min}}$ and $\eta_{\rm{max}}$ is shown in Fig.\,\ref{Fig:EtaRanges}. Existing hadron collider datasets are indicated as horizontal lines at the corresponding center-of-mass energy. Measurements that could be used to improve the modelling for CR background for IACTs should lie within the given pseudorapidity range at the respective energy. The colored boxes represent the pseudorapidity coverage of different detectors that were present during the respective data taking periods. Fig.\,\ref{Fig:EtaRanges} shows that only two detectors have taken data in the relevant pseudorapidity range for CR backgrounds: the LHCf detector \cite{LHCf_TDR} has taken data at $|\eta|>8.4$ in proton-proton collisions at the LHC at energies of $13.6\,\rm{TeV}$, $13\,\rm{TeV}$ and $2.76\,\rm{TeV}$. The RHICf detector \cite{RHICf} has recorded data during proton-proton collisions at RHIC at an energy of $510\,\rm{GeV}$ covering $\eta>6$. The most interesting measurements for the purpose of tuning event generators to improve CR background modelling could come from direct neutral pion reconstruction and energy measurement in the respective pseudorapidity bins. However, there are no such public results from the LHCf or RHICf collaborations yet.  

\section{Outlook for an LHC Proton-Oxygen Collision Run}

While the suggested measurement of highly energetic forward neutral pions in proton-proton collisions at different energies can already provide valuable tuning data that could improve the modelling of the CR background for IACT analyses, one also has to consider that CR induced EAS may originate from a proton-Nitrogen or proton-Oxygen collision, rather than a proton-proton collision. As the modelling of collisions of atomic nuclei includes an even wider set of free parameters in the event generators, the modelling of these interactions is more complex. Furthermore, the event generators have not been tuned yet to these kind of interactions at high energies. This leads to even more pronounced discrepancies in the prediction of neutral pion energy spectra between the event generators for proton-Oxygen collisions than in proton-proton collisions, as shown in Fig.\,\ref{Fig:PionE1TeV} on the right.
To mitigate these modelling discrepancies for CR backgrounds at IACTs, the planned proton-Oxygen collision run at the LHC in 2025 will provide an interesting opportunity for forward neutral pion measurements with the LHCf detector.



\section{Conclusion}
 
Our studies show that CR induced EASs with a highly energetic neutral pion produced at the early shower stages are a main irreducible background to gamma-ray induced EAS analyses with IACTs. It was demonstrated that the available event generators differ strongly in their predictions for this process class. Based on simulations and kinematic arguments we defined a region of interest for proton-proton collision experiments in the $\eta$-$\sqrt{s}$ plane, which relates directly to the CR background. It was shown that out of the considered hadron collider experiments only the LHCf and RHICf detectors covered this area of phase space. Measurements of neutral pions with their data could provide a valuable input for the tuning of the event generators. Furthermore, the event generators could benefit from a tuning based on LHCf data from the planned proton-Oxygen run in 2025.

\bibliographystyle{IEEEtran}
\bibliography{bibliography.bib}

\end{document}